# Electronic structure calculations for PrFe$_4$P$_{12}$ filled skutterudite using Extended Huckel tight-binding method

Donald H. Galvan

*Centro de Ciencias de la Materia Condensada-UNAM, Apartado Postal 2681, CP 22800, Ensenada, B. C., México.*

## ABSTRACT

To gain insight into the electronic properties of PrFe$_4$P$_{12}$ filled skutterudite, band electronic structure calculations, Total and Projected Density of States, Crystal Orbital Overlap Population and Mulliken Population Analysis were performed. The energy bands yield a semi metallic behavior with a direct gap (at Γ) of 0.02 eV. Total and Projected Density of States provided information of the contribution from each orbital of each atom to the total Density of States. Moreover, the bonding strength between some atoms within the unit cell was obtained. Mulliken Population Analysis suggests ionic behavior for this filled skutterudite.

## I. INTRODUCTION

Filled ternary skutterudites[1] are a family of compounds with general structural formula $RT_4X_{12}$ (where R = alkaline earth, rare earth, actinide; T = Fe, Ru, Os; X = pnictogen: P, As, Sb) and crystallize[1,2] in the cubic skutterudite structure (space group *Im3*). A variety of interesting physical properties such as superconductivity[3,4], magnetic order[5], small hybridization gap semiconductivity (also known as "Kondo insulator" behavior)[6], valence fluctuation and heavy fermion behavior[7-10], non-Fermi liquid behavior [11], and metal-insulator transition[12].

Recently, these compounds attracted renewed attention due to their potential application as thermoelectric materials (TE)[13-14]. TE performance is characterized by a dimensionless figure of merit. $ZT = \sigma S^2 T/\kappa$, where $\sigma$ is the electric conductivity, S is the thermopower, and $\kappa$ is the thermal conductivity; ZT up to 1.4 at 600 K has been measured in skutterudites.

Recently, $YbFe_4Sb_{12}$, the first filled skutterudite compound based on a rare earth element heavier than Eu, was synthesized[10,15].

In order to get insight into the extraordinary properties of the filled skutterudite compounds; electronic structure calculations on $PrFe_4P_{12}$ were performed.

The organization of this work is as follows: Sec. II computational details of the calculations. The results and discussion of the electronic properties are presented in Sec. III. Finally, in Sec. IV, we present the conclusion of this work.

## II. COMPUTATIONAL DETAILS

The calculations reported in this work had been carried out by means of the tight-binding method[16] within the extended Huckel[17] framework using YAeHMOP[18] computer package with f orbitals[19]. More details about the mathematical formulation of this method had been describe[20] elsewhere and are omitted here.

The structure of $PrFe_4P_{12}$ used in the present work is illustrated in Fig. 1 from Leithe-Jasper et al.[15] large spheres represents Pr, medium size spheres depicts Fe and smaller spheres represents P atoms respectively. It is characterized by the space group: $Im\text{-}3\text{-}T_h^5$ No. 204, Z=2, and with primitive vector a = 0.7771 nm, V = 0.7679 $nm^3$, with one Pr atom located in the position (2a) 0 0 0, and four Fe atoms located at (8c) 1/4 1/4 1/4, 3/4 3/4 1/4, 3/4 1/4 3/4, 1/4 3/4 1/4; and twelve P atoms located at (24g) 0 y z, 0 -y z, 0 y -z, 0 -y -z, z 0 y, z 0 -y, -z 0 y, -z 0 -y, y z 0, -y z 0, y -z 0, - y -z 0, with y = 0.1593 and z = 0.3365 of π/a. The atomic parameters employed in our calculations are summarized in Table I.

In order to facilitate the analysis in $PrFe_4P_{12}$, it is well known that the unit cell for this compound could be decomposed into $(Fe_4P_{12})^{3-}$ sub units[21], while Pr has a valence of 3+, making the unit cell neutral. Hence, the skutterudite structure is formed by six $P_4$ rings around each Fe atom and can be described by a cube. The center of the cube contains a Fe atom, and each face of a cube contains a quarter of one $P_4$ ring, with the center of the $P_4$ ring located at each cube corner, thus, six different $P_4$ rings provide the octahedral coordination of Fe. The body-center unit for $PrFe_4P_{12}$ is constructed by assembling eight such cubes.

### III. RESULTS AND DISCUSSION

#### A. Electronic Properties

The band structure of the filled skutterudites under study was calculated using 51 K points sampling the First Brillouin Zone (FBZ) as shown in Figs. 2 (a) to (b). Fig. 2 (a) yield an indication of the Valence band (VB) and the Conduction band (CB); the Fermi level ($E_f$) is indicated by a horizontal dotted line. The inset in Fig. 2 depicts the Wigner-Seitz cell for a Base Center Cubic (BCC) configuration spanning from $\Gamma$ (0 0 0) to H (0 $\pi/2a$ 0) to N ($\pi/4a$ $\pi/4a$ 0) to $\Gamma$ ( 0 0 0 ) to P ($\pi/4a$ $\pi/4a$ $\pi/4a$) which generates the first Brillouin zone (FBZ). The most important features in the dispersion diagram for $PrFe_4P_{12}$ is a tendency to form a pseudo gap that appears above the Fermi level (see Fig. 2 (b)), which is a characteristic of skutterudites, especially when a single phosphorous p-band crosses the Fermi level. The crossing band is indeed pushed down by the repulsion of Pr f-resonance states.

In order to study more closely the bands around the Fermi level, these bands have been reploted on an expanded energy scale in Fig. 2 (b). A direct mini gap measured at $\Gamma$, between the highest and two horizontal lines indicate lowest bands in the VB and CB. The $E_g$ obtained was of the order of 0.02 eV.

Performing a more careful analysis to the two bands in the vicinity of the Fermi level, see Fig. 2 (b), these two bands form a hybridize band. Note at $\Gamma$, the highest occupied band forming the VB has a parabolic shape, which implies a regular effective mass, on the other hand, the lowest unoccupied band from the CB has a more pronounced shape, implying an effective mass bigger than the former one. The combination of these two bands, with different effective mass in the vicinity of the Fermi level, is what provides the utility of $PrFe_4P_{12}$ as a thermoelectric material.

Summarizes in Table II are the values obtained for the energy gap ($E_g$)(direct and indirect), the VB and CB for the ternary skutterudites $LaFe_4P_{12}$, $CeFe_4Sb_{12}$, $CeFe_4P_{12}$ and $YbFe_4Sb_{12}$, as well as values derived from other methods of calculation and from the experiment for each compound, when available.

For the ternary skutterudites, Nordstrom et al.[22] performed a density functional calculation within the local-density approximation (LDA) obtaining indirect gaps in $CeFe_4P_{12}$ and $CeFe_4Sb_{12}$, of 0.37 and 0.10 eV respectively, while Dordevic et al.[23] reported an experimental gap of 15 meV.

Jung et al.[21] using extended Huckel tight-binding method calculated the value gap of 2.7 eV for the indirect gap of $LaFe_4P_{12}$. While Fornari et al.[24] performed calculations within the framework of the density functional theory using the general potential linearized augumented plane wave (LAPW) method and obtained 98 meV for an indirect gap measured at Γ. On the other hand, Meisner et al.[25] found for $LaFe_4P_{12}$ a metallic behavior using transport measurements, also found superconductivity behavior with a transition temperature of 4.2 K. Galvan et al.[26] using tight-binding method within the Extended Huckel framework using YAeHMOP computer package with f-orbital for Yb (U or Th), reported no evidence for the existence of a mini gap in $YbFe_4Sb_{12}$, $UFe_4P_{12}$ and $ThFe_4P_{12}$ respectively.

Our calculations for $PrFe_4P_{12}$ yield a mini gap of the order of 0.02 eV, as shown in Fig. 2(b). Furthermore, it is good to underline, that only a single band from the VB crosses the Fermi energy. This behavior has been observed in filled skutterudites, implying the semi metallic character of this compound. The difference obtained between the experimental values and the presence of a mini gap in our calculation is a good indication that this compound could be used as an effective thermoelectric material. On the other hand, the absence of a mini gap in $YbFe_4Sb_{12}$, $UFe_4P_{12}$ and $ThFe_4P_{12}$ reported by our group[26] could be attributed to the omission from our calculations of either electronic correlations or spin-orbit coupling, which are necessary in these kinds of materials where many electrons are involved.

Unfortunately, up today no experimental and theoretical data are available in order to compare with.

For the width of the VB and CB, we obtained the values 9.35 and 29.37 eV, respectively.

### B. Density of States

The total and projected density of states (PDOS) for the filled skutterudite $PrFe_4P_{12}$ are depicted in Figs. 3 (a) - (g) for Pr (f, d, p and s orbital), Fe (d, p and s orbital) and P (p and s orbital). For Pr the p and s orbital were not included in the energy window because they did not contribute appreciably. In these figures, energy in eV (vertical axis) is plotted *vs* % contribution (horizontal axis); the solid line represents the total DOS, while the dotted and hatched lines are the selected projected DOS for each orbital. The horizontal dotted line indicates the Fermi level $E_f$. It is good to stress that each orbital of each atom had been projected separately in order to find the contributions attributed from each atom to the total DOS in the specific energy range selected.

One of the most important properties about this compound is the possibility of TE applications in industry. Henceforth, we analyze the orbital in the vicinity of the Fermi level, from -9.5 to -12.2 eV. The main contributions to this band are from Pr f-orbital (50 %), Fe d-orbital (2.5 %) and P p-orbital (3 %). Also, notice that the Fermi levels is located almost at the middle of Pr f-orbital, an indication that hybridization between this orbital with Fe d- orbital and P p-orbital had occurred.

Dordevic et al.[23] claim similar results in $YbFe_4P_{12}$, while Galvan et al.[26] obtained similar results in $YbFe_4Sb_{12}$, $UFe_4P_{12}$ and $ThFe_4P_{12}$.

In order to provide more information related to the hybridization occurred between Pr f-orbital, Fe d-orbital and P p-orbital, Table III is provided. This table provides the amount of overlap (in %) of the respective orbital from Fe and P with Pr f-orbital. Note that Fe d, and p orbital contributes with 8 and 1 % respectively, while P p-orbital contributes 3 %. It is good to point out that these overlap of several orbitals in the vicinity of the Fermi level yield the expected hybridization as occurred in other ternary skutterudites.

What kind of information is provided from the location of Pr f-orbital whether they are located above, in between or bellow the Fermi level?

From our experience obtained while performing similar calculations on the filled skutterudites such as $YbFe_4Sb_{12}$, $UFe_4P_{12}$ and $ThFe_4P_{12}$ and performing a careful analysis on Tables III, notice that the aforementioned hybridization between Pr f-orbitals with Fe d, p orbitals and P p-orbitals, the values obtained are not very big, implying that the hybridization occurred is not very strong, which provides an indication that $PrFe_4P_{12}$ is a mild semi metallic compound.

From the thermoelectric point of view, the contribution from each atom to the hybridize band, the stronger the contribution to the band the more likely that the compound under study could be considered a likely candidate as a TE material.

### C. Crystal Orbital Overlap Population (COOP)

In order to obtain information about bonding between some atoms in the unitary cell, Crystal Overlap Population Analysis (COOP) were performed. The integral of the COOP curves provides the total overlap population, which is not identical to the bond order, but scales like it [27].

The amplitude of these curves depends on the number of states in the specific energy interval, the magnitude of the coupling overlap, and the size of the coefficients in the molecular orbital under consideration.

Energy in eV *vs* positive (bonding) or negative (antibonding) contributions (in %) from each orbital to each band is provided.

In order to perform the analysis, the following atoms were selected: $Pr_{17}$-$Fe_1$ (3.4092 A), $Pr_{17}$-$P_5$ (3.4092 A), $Fe_1$-$P_5$ (2.5609 A), $Fe_3$-$Fe_4$ (6.4750 A) and $P_5$-$P_6$ (2.9175 A). The number in parentheses provides the distance in between the selected atoms. Notice that depending on the selected atoms, the corresponding bonding or antibonding and also the magnitude of the overlap are provided. An indication of the strength on the specific bond between atoms is provided which depends on the distance in between atoms and the electronegativity of each individual

### D. Mulliken Population Analysis

Mulliken population analysis is defined indices to quantitatively locate electronic charge around an atom, and its bonding or anti bonding nature[28].

As we mentioned in Sec. II, the sub unit $(Fe_4P_{12})^{3-}$ (pnictogen four member ring), Fe is an octahedral coordination with P. Hence, the orbitals that contribute to form the hybridize band in the vicinity of the Fermi level are Fe d-orbital. These five orbitals separate into $t_{2g}$ ($d_{xy}$, $d_{xz}$, $d_{yz}$) and $e_g$ ($d_{x2-y2}$, $d_{z2}$) symmetric bands. Due to the crystal field splitting, being $t_{2g}$ block band lower in energy than $e_g$ block band, hence it is expected that these orbital will electronically be filled in that order.

In this research, the d-orbitals are filled in the following order $d_{yz}$ (1.917), $d_{xy}$ (1.909), $d_{xz}$ (1.908), $d_{z2}$ (1.878) and $d_{x2-y2}$ (1.856). The number in parenthesis provides the electronic occupation being 2 the maximum number of electrons in a single orbital.

In P the orbitals that are filled first are s-orbitals (1.235) and then $p_z$ (1.044), $p_x$ (1.04) and last $p_y$ (0.872). Notice that both orbitals are more than half-filled.

Furthermore, Pr f-orbitals are filled as follows: $f_{yz2}$ (1.385), $f_{xyz}$ (1.328), $f_{z3}$ (1.315). These orbitals are more than half-filled. The rest of the f-orbitals are filled in the subsequent order: $f_{xz2}$, $f_{z(x2-y2)}$, $f_{x(x2-3y2)}$ and $f_{y(3x2-y2)}$. These orbitals are less than half-filled.

The fact that some of these orbitals are completely filled, and some other are more than half-filled, provides indication that a none uniform charge distribution (polarization) exist between some atoms in the unit cell. From this analysis, it is possible to infer that some kind of ionic behavior exist in $PrFe_4P_{12}$.

## IV. CONCLUSIONS

The following conclusions can be drawn from the calculations described herein for $PrFe_4P_{12}$:

- The energy bands indicate that this compound is a semi metal, with a clear distinction between CB and VB. A direct gap measured at Γ yield 0.02 eV, while Dordevic et al.[29] obtained 13 meV for $YbFe_4Sb_{12}$.

- The character of the bands was identified as s, p, and d and f electrons for Pr, s, p and d for Fe, and s, p for P respectively.

- The important features of a likely pseudogap created by the P p-orbital that crosses the Fermi level with the interaction of Pr f-resonance states provide indication that this material is a likely candidate for TE applications.

- From the total and projected DOS, it was possible to identify the contributions from each orbital from each atom, in $PrFe_4P_{12}$.

- The band from - 9.5 to -12.2 eV, is mainly formed from the contributions from Pr f-resonance states with an admixture of small contributions of P p-orbital and Fe d-orbital. Moreover, hybridization occurs within this band due to the overlap of Fe d and p orbital, P p orbital with Pr f-orbital, as summarized in Table IV. This hybridization is consistent with the predictions of Dordevic et al.[29] in $YbFe_4Sb_{12}$.

  A similar feature has been reported in heavy fermion compounds such as $URu_2Si_2$ where a partial gap forms at the Fermi surface in association with a spin-density wave ground state[30].

- The COOP analysis provided information related to the bonding strength between some selected atoms within the unit cell and their character (bonding or antibonding).

- Mulliken population analysis yield information related to the ionic behavior in $PrFe_4P_{12}$.

## ACKNOWLEDGMENTS

D. H. Galván thanks acknowledges A. Aparicio for technical support, DGAPA - UNAM and UNAM-CREY under Grant No. SC-004696 for providing supports.

# REFERENCES

#Author to whom correspondence should be addressed: Centro de Ciencias de la Materia Condensada-UNAM, P. O. Box 439036, San Ysidro, CA 92143, USA.

**FIGURE CAPTIONS**

Figure 1 Unit cell for $PrFe_4P_{12}$. Large spheres represent Pr atoms, medium size spheres correspond to Fe atoms and small spheres are P atoms.

Figure 2 (a) - (b) Band structure for $PrFe_4P_{12}$. The Fermi energy ($E_f$) is indicated by a dotted line. The inset depicts the Wigner-Seitz primitive cell for a BCC configuration, which generates the FBZ, selected.

Figures 3 (a) - (g) Total and projected Density of States (PDOS) for Pr, Fe and P. The dotted and hatched lines emphasize the contributions to the total DOS from each atom.

Figures 4 (a) - (f) Depict the Crystal Orbital Overlap Population (COOP) for $PrFe_4P_{12}$.

TABLE I. Atomic parameters used in the Extended Huckel tight-binding calculations, $H_{ii}$ (eV) and ς (Valence orbital ionization potential and exponent of Slater type orbitals). The d-orbitals for Fe and Pr are given as a linear combination of two Slater type orbitals. Each exponent is followed by a weighting coefficient in parentheses. A modified Wolfsberg-Helmholtz formula was used to calculate $H_{ij}$[31].

| Atom | Orbital | $H_{ii}$ | $ς_{i1}$ | $C_1$ | $ς_{i2}$ | $C_2$ |
|---|---|---|---|---|---|---|
| Fe | 4s | -9.10 | 1.9 | | | |
|  | 4p | -5.32 | 1.9 | | | |
|  | 3d | -12.60 | 5.32 | (0.5505) | 2.00 | (0.6260) |
| P | 3s | -18.60 | 1.75 | | | |
|  | 3p | -14.00 | 1.30 | | | |
| Pr | 6s | -7.42 | 1.40 | | | |
|  | 6p | -4.65 | 1.40 | | | |
|  | 5d | -8.08 | 2.753 | (0.7187) | 1.267 | (0.4449) |
|  | 4f | -11.28 | 6.907 | (0.7354) | 2.639 | (0.4597) |

TABLE II. Values (in eV) for the energy gap $E_g$ and width of the Valence and Conduction bands, respectively.

|  | $LaFe_4P_{12}$ | $CeFe_4P_{12}$ | $CeFe_4Sb_{12}$ | $YbFe_4Sb_{12}$ | $UFe_4P_{12}$ | $ThFe_4P_{12}$ | **$PrFe_4P_{12}$** |
|---|---|---|---|---|---|---|---|
| $E_g$ Indirect Direct | 2.7[a] 0.0098[c] | 0.37[b] | 0.10[b] | None[e] 0.0013[d] | None[e] | None[e] | **0.02 (at Γ)** |
| Width (VB) |  |  |  | 4.89[e] | 8.23[e] | 9.7[e] | **9.35** |
| Width (CB) |  |  |  | 11.40[e] | 28.18[e] | 28.3[e] | **29.37** |

[a] D. Jung, M. H. Whangbo and S. Alvarez, Inorg. Chem., **29,** 2252 (1990). Tight-binding method within the Extended Huckel framework.

[b] L. Nordstrom and D. L. Singh, Phys. Rev. B **53,** 1103 (1996). Density Functional calculations within the Local Density Approximation (LDA).

[c] M. Fornari, D. J. Singh, Phys. Rev. B **59,** 9722 (1999). Density Functional Theory using the general potential linearized augumented plane wave (LAPW) method.

[d] S. V. Dordevic, N. R. Dilley, M. B. Maple, and D. N. Basov, Phys. Rev. Lett., **86,** 684 (2001).

[e] D. H. Galvan, N. R. Dilley, M. B. Maple, A. Posada Amarillas, and A. Reyes Serrato, to be published.

TABLE III. Percentage of the overlap of the respective orbitals of Fe and P with Pr f-orbitals, in the energy band between -9.5 to -12.2 eV, where hybridization occurs.

| Fe | Fe | Fe | Pr | Pr | Pr | Pr | P | P |
|----|----|----|----|----|----|----|---|---|
| d | p | s | f | d | p | s | p | s |
| 8.0 | 1.0 | | | | | | 3.0 | |

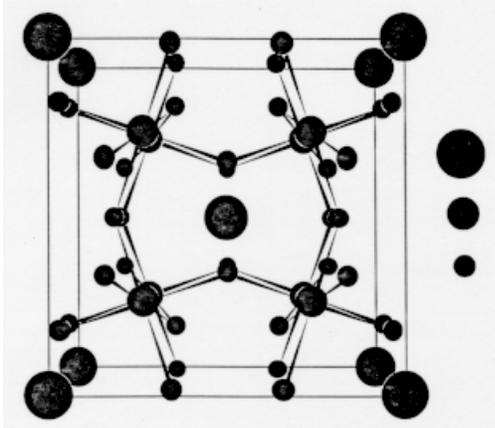

Figure 1.

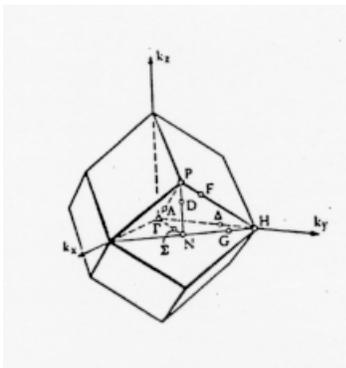

(Inset)

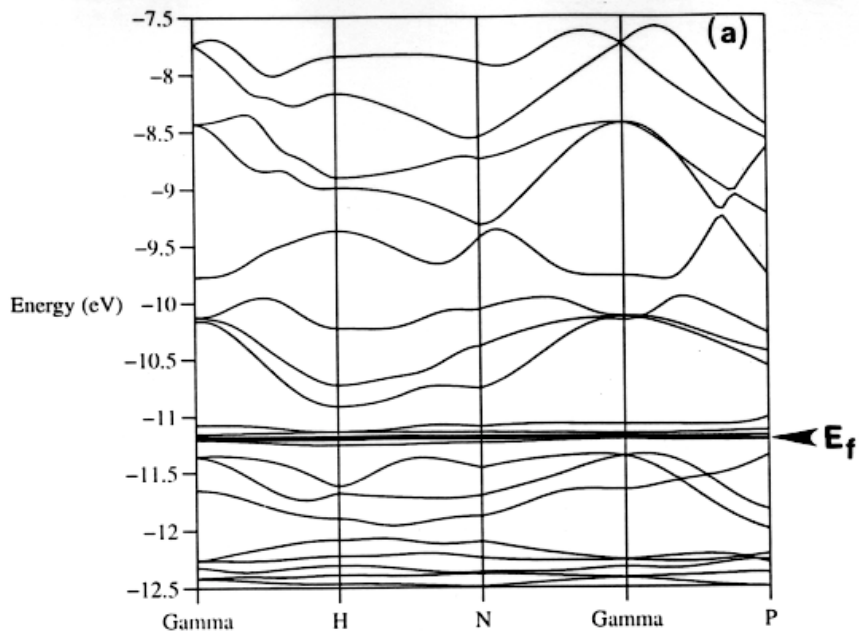

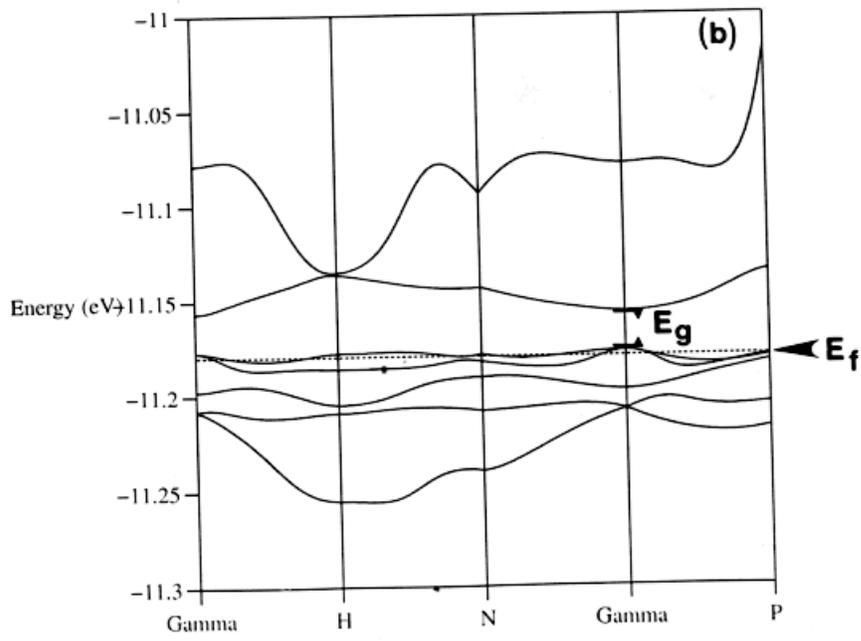

Figures 2

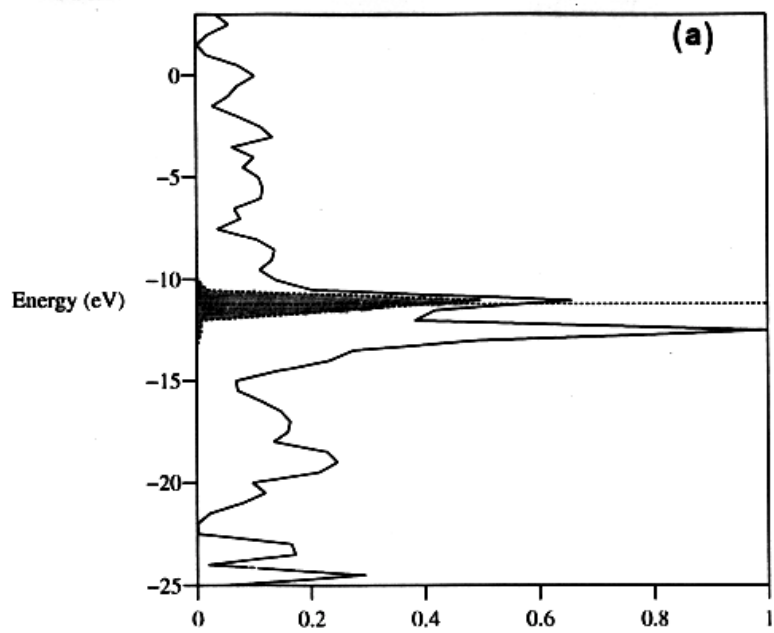

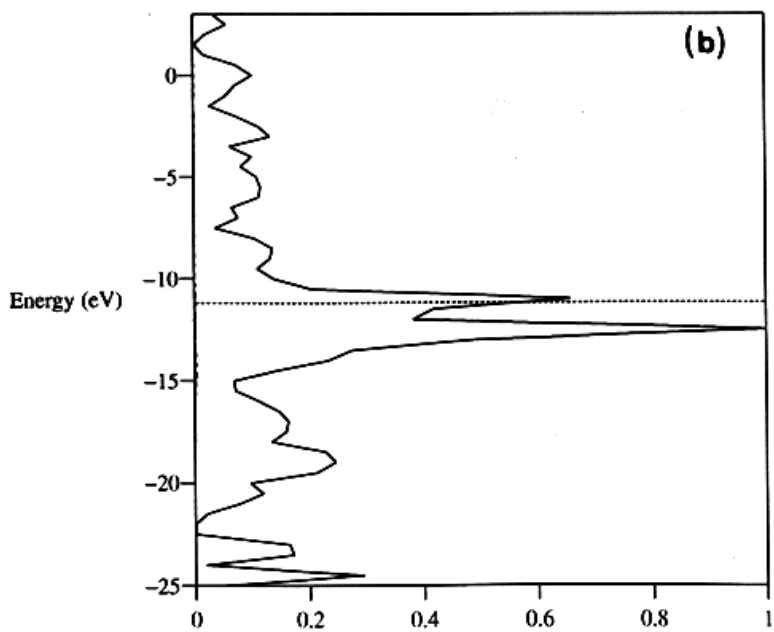

Figures 3

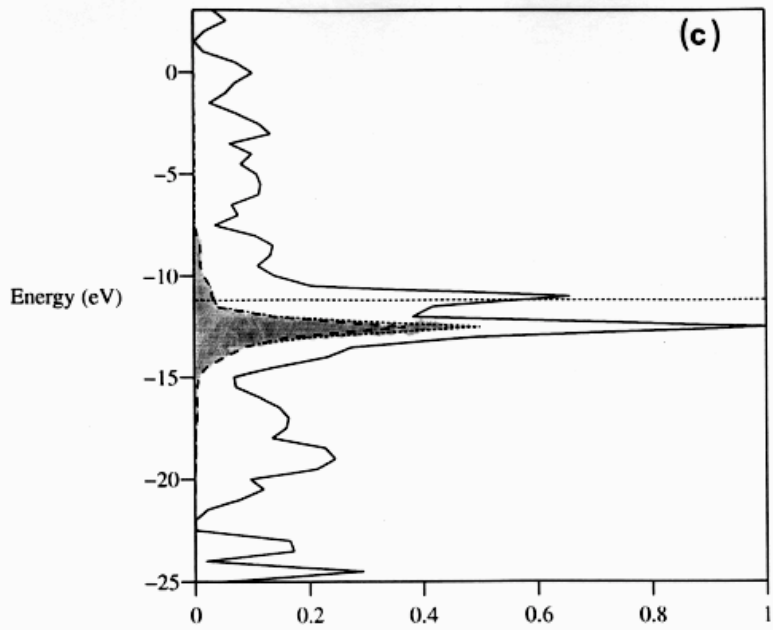

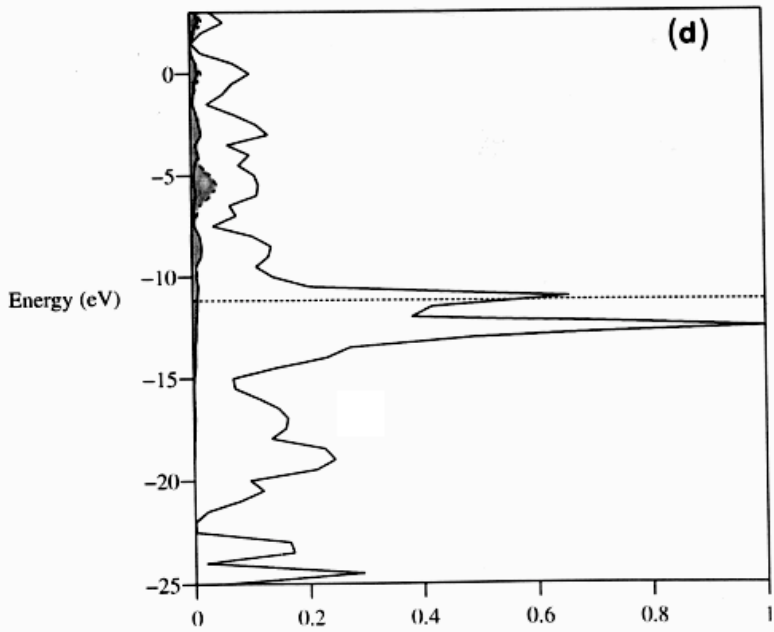

Fig.

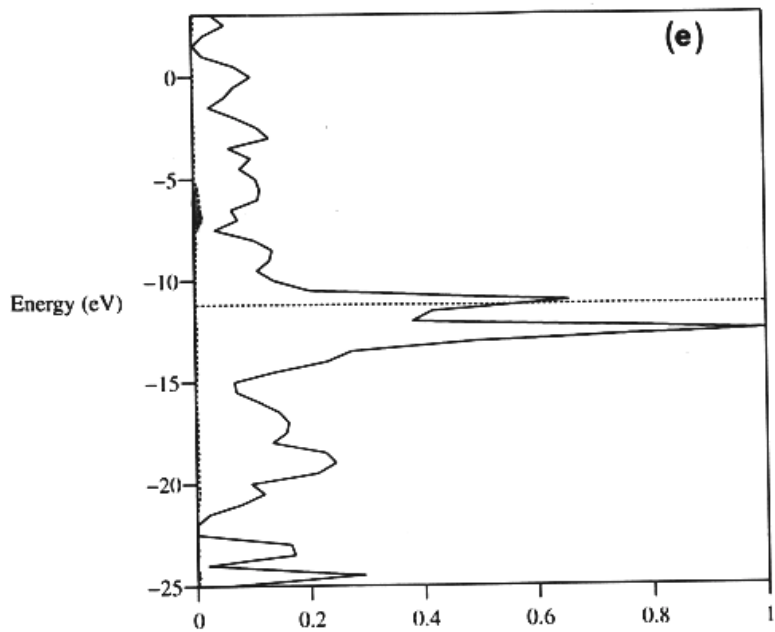

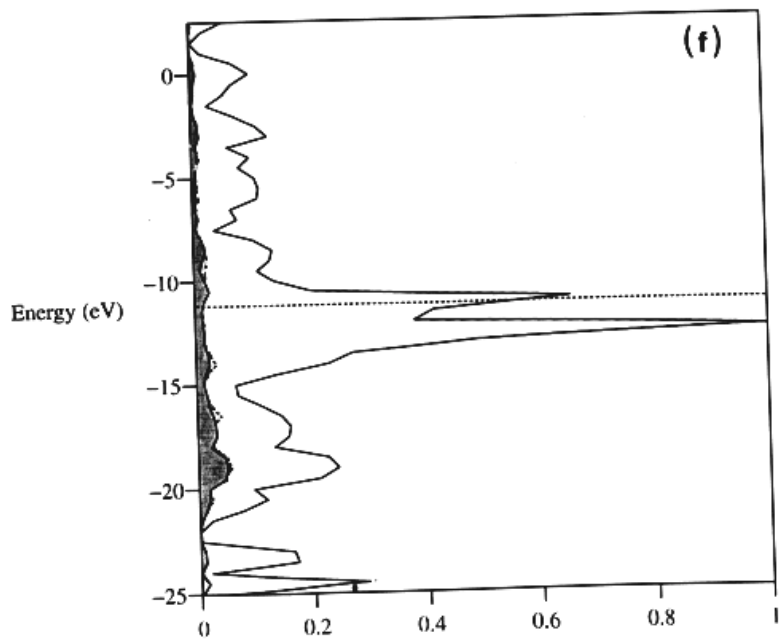

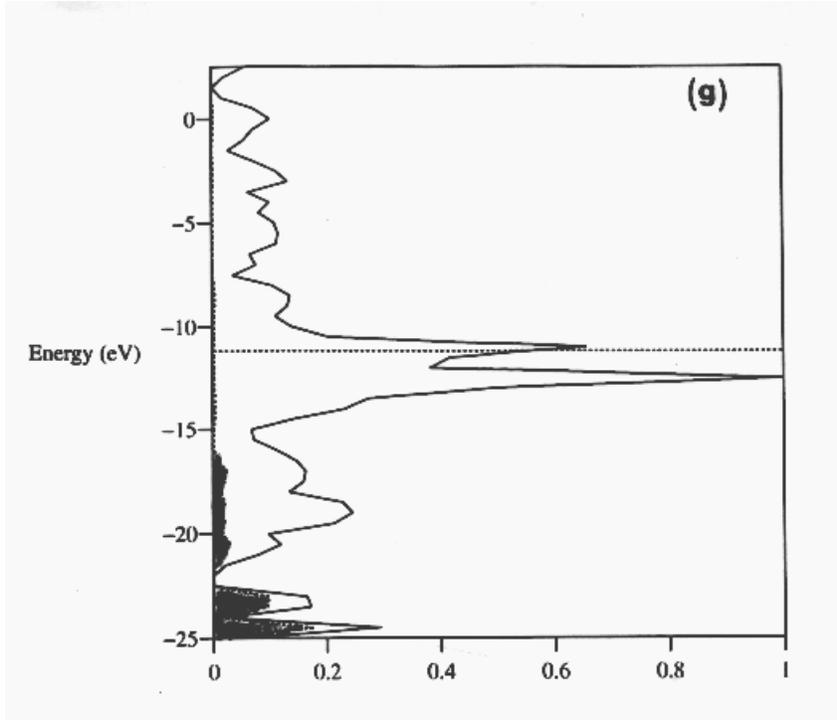

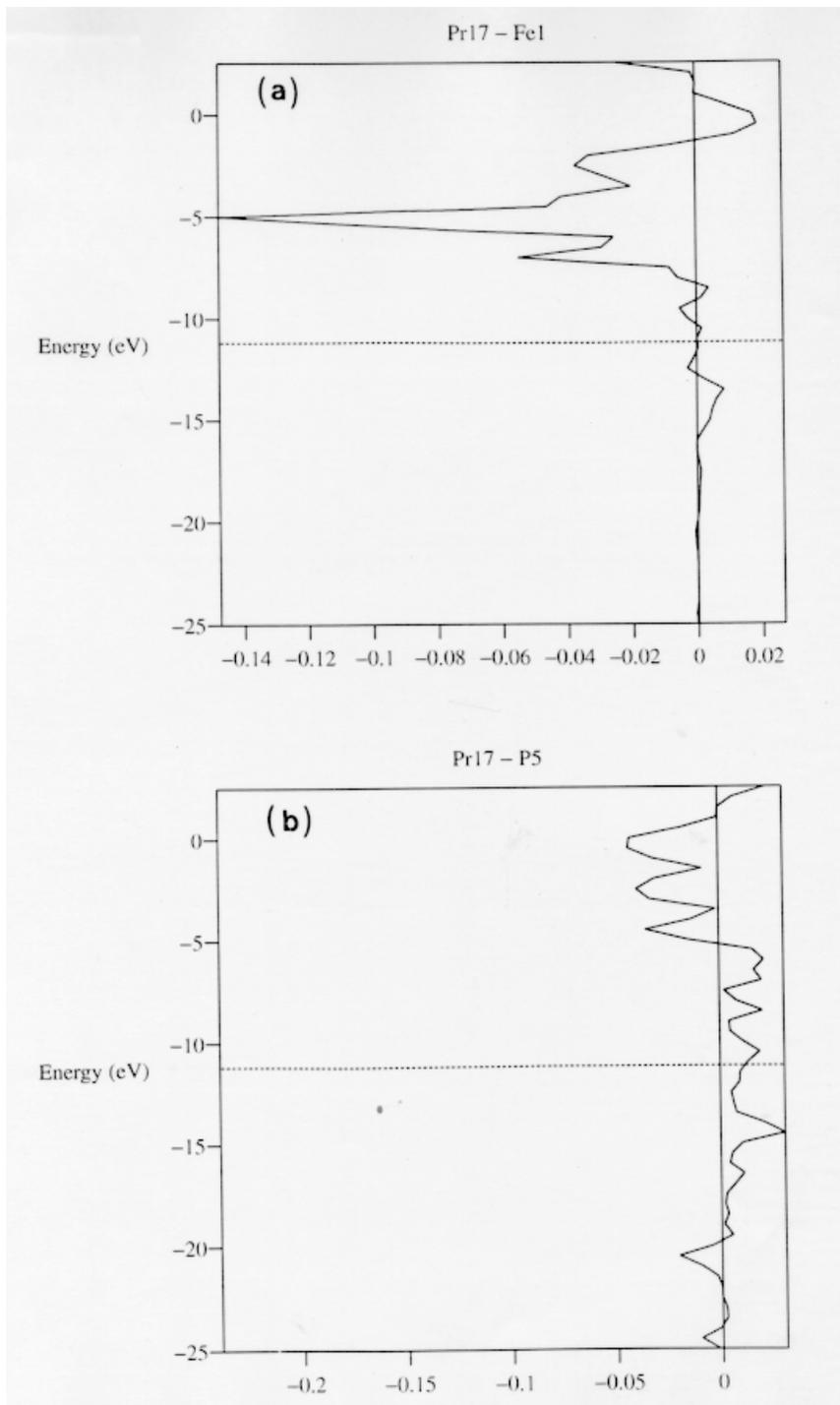

Figures 4

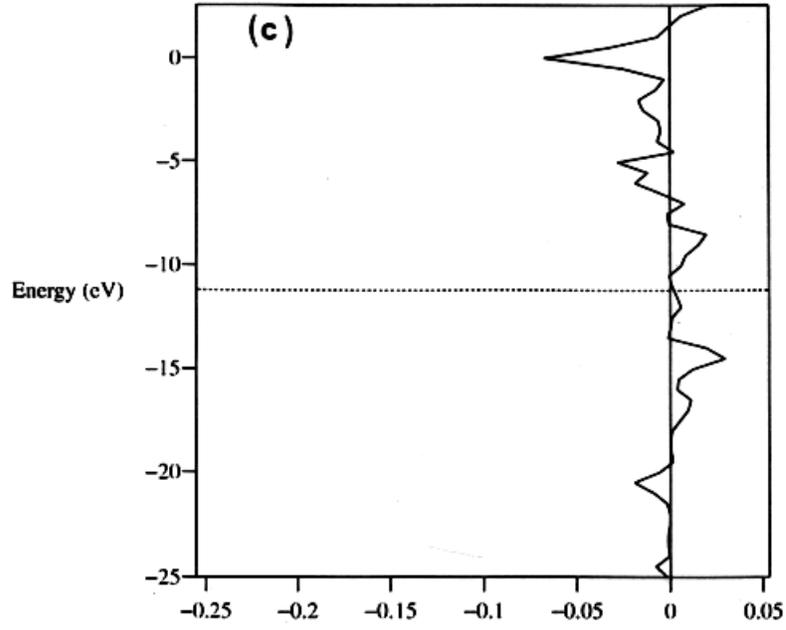

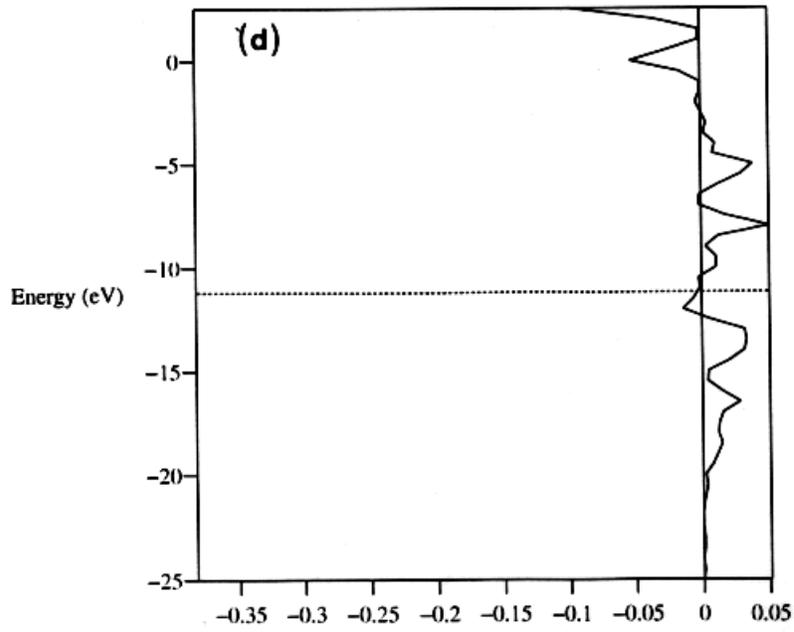

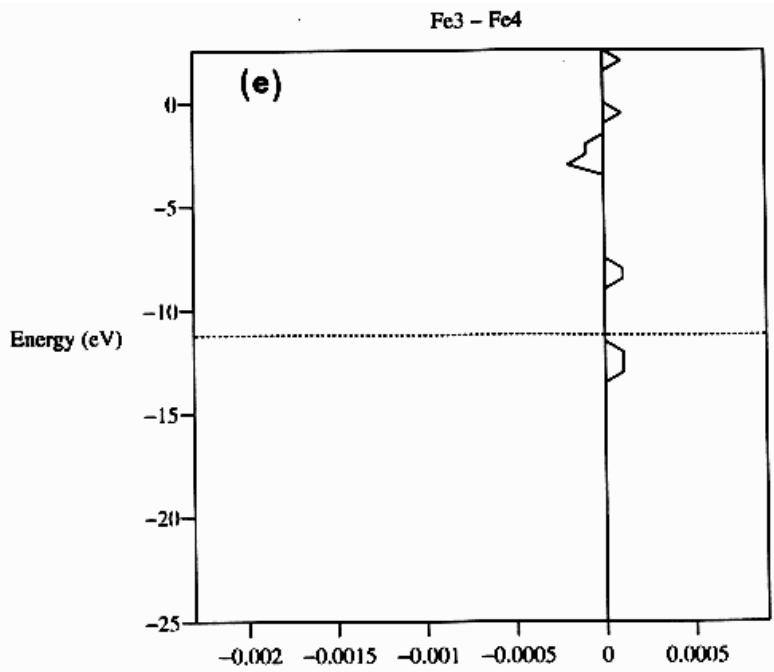

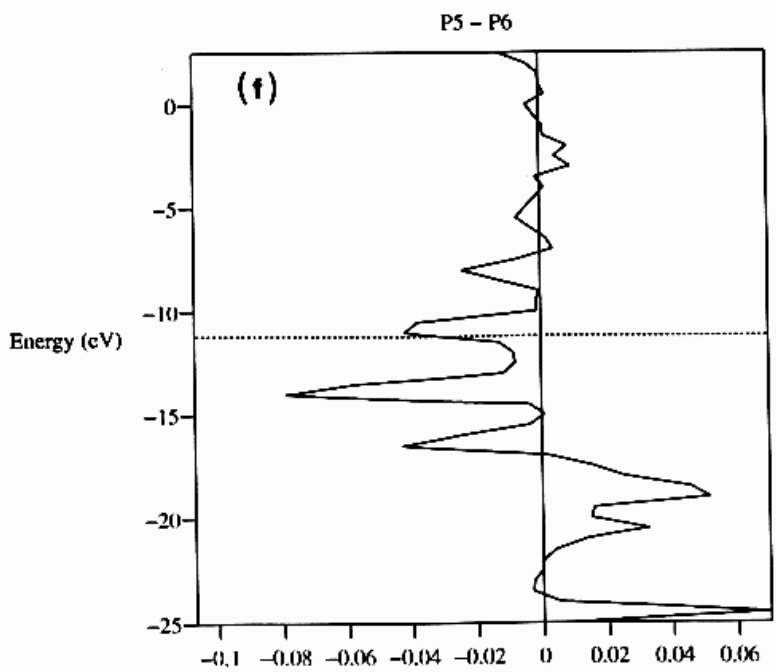